# Presentation Attack detection using Wavelet Transform and Deep Residual Neural Net

Prosenjit Chatterjee[1], Alex Yalchin[2], Joseph Shelton[3], Kaushik Roy[4], Xiaohong Yuan[5] and Kossi D. Edoh[6]

[1,4,5]Dept. of Computer Science, North Carolina A&T State University, USA
[2]Dept. of Computer Science, Elon University, North Carolina, USA
[3]Dept. of Engineering and Computer Science, Virginia State University, USA
[6]Dept. of Mathematics, North Carolina A&T State University, USA
[1]pchatterjee@aggies.ncat.edu, [2]ayalcin@elon.edu, [3]jshelton@vsu.edu, [4]kroy@ncat.com, [5]xhyuan@ncat.edu, [6]kdedoh@ncat.edu

**Abstract.** Biometric authentication is becoming more prevalent for secured authentication systems. However, the biometric substances can be deceived by the imposters in several ways. Among other imposter attacks, print attacks, mask-attacks, and replay-attacks fall under the presentation attack category. The biometric images, especially iris and face, are vulnerable to different presentation attacks. This research applies deep learning approaches to mitigate the presentation attacks in a biometric access control system. Our contribution in this paper is two-fold: First, we applied the wavelet transform to extract the features from the biometric images. Second, we modified the deep residual neural net and applied it on the spoof datasets in an attempt to detect the presentation attacks. This research applied the proposed approach on biometric spoof datasets, namely ATVS, CASIA two class, and CASIA cropped image sets. The datasets used in this research contain images that are captured in both a controlled and uncontrolled environment along with different resolution and sizes. We obtained the best accuracy of 93% on the ATVS Iris datasets. For CASIA two class and CASIA cropped datasets, we achieved test accuracies of 91% and 82%, respectively.

## 1 Introduction

Biometric authentication uses an individual's identity for access control and has been widely implemented for controlling the secured gateway of the member's login access. Several organizations validate their members' access through biometric-enabled surveillance and security systems. The earlier biometric authentication techniques utilized physiological traits such as fingerprint, face, iris, periocular region, voice, heart rate, and body mass. Biometric authentication systems have evolved further to the individual behavioral pattern such as touch pattern on electronic devices, keystroke dynamics, etc., to distinguish real and fake individual identity. It is evident that the human iris and facial biometric image samples are vulnerable to different types of presentation attacks.

Though the presentation attack falls under the 'hacking attack' category, it is also referred to as a replay attack. There are several other vulnerable points that the hackers exploit to compromise a biometric-based authentication system. Fake human faces can be created through 3D printing devices, by the use of 3D Mask, or presenting an identical twin, or similar looking individual, to deceive an authentication system. Additionally, the human iris can be copied using textured contact lenses to deceive an iris-based authentication system. A key research focus is the attempt to build a robust classification technique to identify the smallest deviation on real and fake iris and facial images in order to detect presentation attacks.

Recently, deep convolutional neural networks have been used to mitigate presentation attack [7 - 9]; however, most of them require huge computational time to train and classify real and fake image samples. To counter presentation attacks, a high-resolution image set is required. However, high-quality images increase computational complexity during training, validation, and classification. To address this issue, we apply a deep learning approach that can detect spoofing attacks with less computational effort and time.

In this paper, we apply the Wavelet Transform [6] on image datasets to extract features. Once the feature extraction is done, we feed the extracted features to a modified Residual Neural Net 'modified-ResNet' inspired by the Residual Neural Net (ResNet) [1] for accurate classification. The ResNet has superior features such as batch normalization, parameter optimization, and reduced error through skip layer connection techniques implemented, with surprisingly less computational complexity. We observe a significant improvement in classification time with high accuracy, to distinguish real and fake images.

The rest of the paper is organized as follows. Section 2 discusses our related work. Section 3 describes the proposed methodology design and a high-level flow diagram, the architecture of ResNet and its implementation. Section 4 discusses the datasets used for this research effort, Section 5 shows the experimental results and discussion, and Section 6 provides our conclusion and future work.

## 2   Related Work

A reliable biometric recognition system has long been a prominent goal to mitigate presentation attacks in a wide range. Recently, deep learning techniques have been used for presentation attack detection (PAD) and have become instrumental to many secured organizations where biometric authentication is mandatory. Some specific well-known spoofing instruments, like silicon masks, are widely used by attackers. Yang et. al [7] used all the testing protocols and implemented facial localization, spatial augmentation, and temporal augmentation, to diverse feature learning for the deep convolutional neural network (deep CNN). They also experimented with texture based, motion-based, 3D Shape-based detection techniques to identify genuine and fake individuals [6]. Menotti. et. al [8] performed research on deep representation for detecting presentation attacks in different biometric substances. Manjani et. al [9] proposed a multilevel deep dictionary learning-based PAD algorithm that can discriminate among different types of attacks. However, the biometric recognition system using the deep CNN technique is completely dependent on recognizing the pattern of test objects with the previously learned training objects. A successful match includes accurate pattern matching of the feature-sets from the test object to the already learned training object. The vital biometric information of an individual, such as faces and iris, are essential for the training. For "mug shot" images of faces taken at least one year apart, even the best current algorithms can have error rates of 43%–50% [10 - 12]. Henceforth, face and iris images of an individual, when implemented altogether on a biometric recognition system, is proven to be authentic amongst other biometric substances and many researchers agreed on that.

Several researchers worked on the face and iris recognition system and proposed diverse methods to extract features in detail [13]. Feature extraction through Wavelet involves losses at the edges after the Label 1 decomposition. We investigated the Biorthogonal wavelet transform [6], the Discrete Wavelet Transform (DWT) [6] and 2D-Gabor wavelets [14] and their inverse form. From the experiment, we found that DWT [6] works better with images in the matrix format ($n$ x $m$) and can convolve in either direction, which facilitates feature augmentation including spatial and temporal augmentation during the deep learning training phase.

After thorough analysis and experiments, we found out that Residual Neural Network (RNN), especially ResNet [1], is the most effective deep learning approach for training and validation process. RNN utilizes single layer skip connection techniques during learning on the internal convolutional layers and avoids the vanishing gradients issues and optimize the huge parameters efficiently. Additionally, the effective single layer skipping makes the network less complex during the initial training phase, and towards the end of the training, all layers get expanded for detailed level learning.

## 3    Proposed Methodology

In this research, we used the discrete wavelet transform (DWT) [6] and its inverse form to elicit feature from the face and iris images. We then implemented a 'modified ResNet', inspired by the ResNet [1], in an attempt to mitigate the presentation attacks. We trained, validated and tested the ResNet model for the images captured under controlled and uncontrolled environments. At the start, we implemented the discrete wavelet transform (DWT) [6] and its inverse form (IDWT) [6] that help the feature extraction from the face and iris images. The extracted features are then fed into our proposed 'modified ResNet' for accurate classification. The high-level flow diagram of our pro-

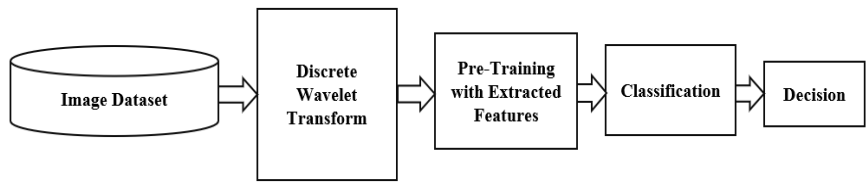

Fig1. High Level Architecture of Our Proposed Methodology

posed methodology is shown in Fig1. To the best of our knowledge, no research work, which has used the combination of wavelets and 'modified ResNet' for PAD, has been reported. In this effort, we used DWT for feature extraction and decomposition and the inverse form of it for reconstruction.

The discrete function is represented as a weighted sum in the space spanned by the bases 𝜑 and 𝜓:

$$x[m] = \frac{1}{\sqrt{N}} \sum_k W_\varphi[j_0, k]\varphi_{j_0,k}[m] + \frac{1}{\sqrt{N}} \sum_{j=j_0}^{\infty} \sum_k W_\psi[j,k]\psi_{j,k}[m], \quad (m = 0, \cdots, N-1) \quad \ldots (1)$$

The inverse wavelet transform, where the summation over *j* is for different scale levels and the sum over *k* is for different conversions in each scale level, and the coefficients (weights) are forecasts of the function onto each of the basis functions:

$$W_\varphi[j_0, k] = (\mathbf{x}, \varphi_{j_0,k}) = \frac{1}{\sqrt{N}} \sum_{m=0}^{N-1} x[m]\varphi_{j_0,k}[m] \quad - \quad V(k) \quad \ldots (2)$$

$$W_\psi[j, k] = (\mathbf{x}, \psi_{j,k}) = \frac{1}{\sqrt{N}} \sum_{m=0}^{N-1} x[m]\psi_{j,k}[m] \quad - \quad V(k) \ \& \ V(j > j_0) \quad \ldots (3)$$

where $W_\varphi[j,k]$ is called the approximation coefficient and $W_\psi[j,k]$ is called the detail coefficient.

We limit our wavelet decomposition to Label-1, to prevent data loss at edges. The feature extraction, decomposition, and reconstruction mechanism we have implemented for our research is shown in Fig2.

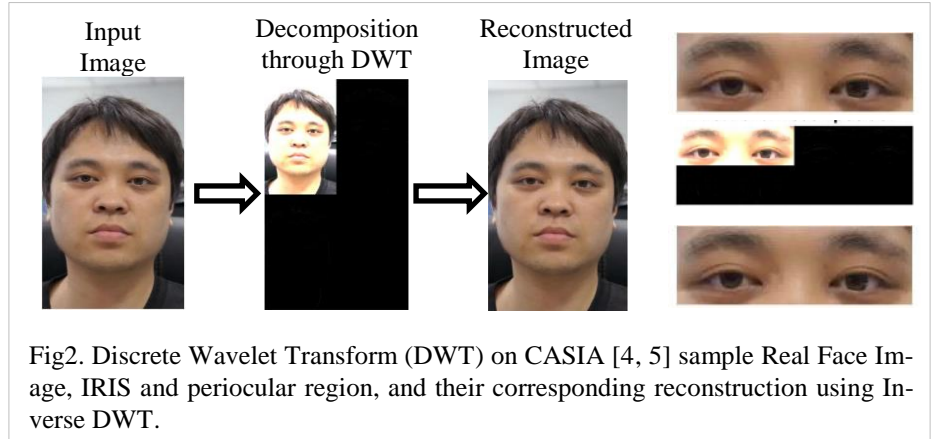

Fig2. Discrete Wavelet Transform (DWT) on CASIA [4, 5] sample Real Face Image, IRIS and periocular region, and their corresponding reconstruction using Inverse DWT.

To prevent presentation attack in a wide range and to save training and execution time, we designed a 'modified ResNet' inspired by the ResNet [1]. Kaiming He et al. [1] proposed and implemented an 18 layer and 34-layer ResNet respectively. They experimented on their proposed skip connection techniques implementations and handled images through the deep CNN structure and showed the advantages they achieved based on computational complexities and low error rate.

In comparison, our 'modified ResNet' is quite lighter than the ResNet [1], as our 'modified ResNet' has a total of 32 discrete convolution 2D layers, 2 Max Pooling 2D layers and 1 Fully Connected (FC) dense layer, shown in Fig.3(a) and Fig.3(b). Our

'modified ResNet' has 5 important building blocks of convolution layers and their distribution maintaining the single layer skip-connection techniques are shown in Fig.3(c)

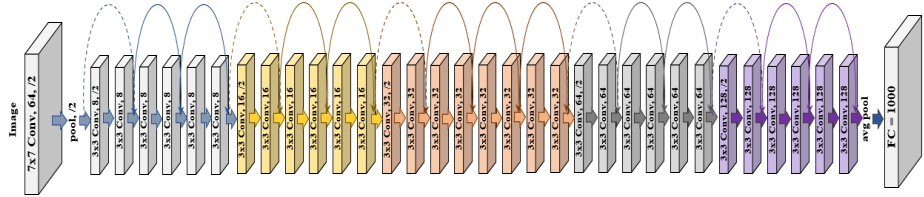

Fig3 (a). Modified ResNet Framework Structure.

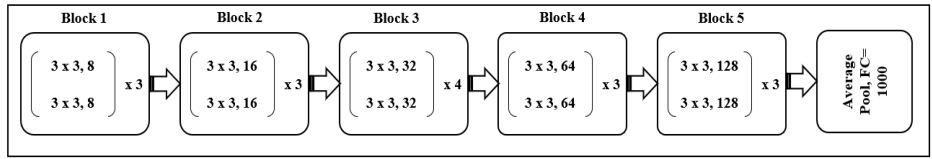

Fig3 (b). Modified ResNet Framework Structure in blocks and their convolution layer distribution.

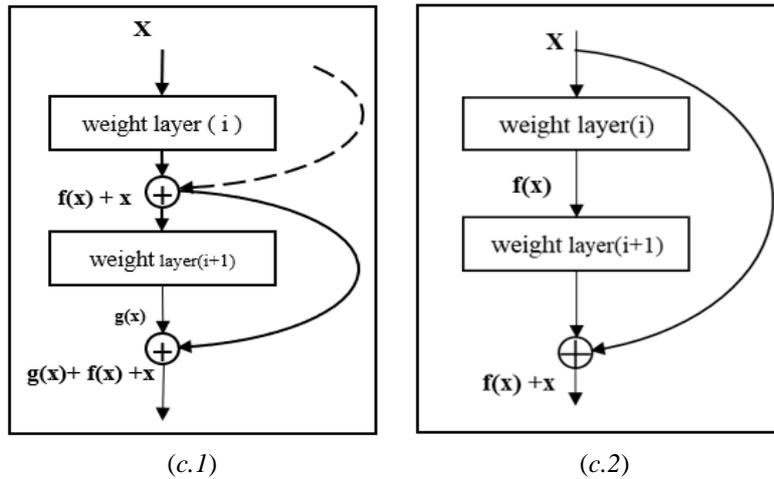

              (c.1)                                (c.2)
Fig 3(c). Residual Neural Net single layer skip sequential function implemented.
    *(c.1) – represents single layer skip connection techniques implemented.*
    *(c.2) – represents double layer skip connection techniques implemented.*

## 4 Datasets Used

Our research scope and results are monitored and captured based on the three different categories of image datasets: ATVS [2, 3], CASIA [4, 5], and CASIA-cropped [4, 5]. ATVS [2, 3] is an iris image dataset that consists of periocular images of human eyes. The periocular region includes iris, eyebrows, eyelids, and eye leaf. The datasets contain fifty subjects. Subjects had both eyes photographed four times each of two different sessions. Each image was then undertaken through gray-scaled printing and successive scanning for fake image generation. Each user contains 32 images, 800 per class (real and fake), and a total of 1600 with a uniform resolution of 640x480.

CASIA [4, 5] dataset contains both the high resolution still images and video clips. There contain fifty subjects and four classes. Every class contains one landscape-style video and one portrait-style video. The four classes are real subjects, "cut photo" attacks (printed photo of subject with eyeholes cut out, real user positioned behind photo to fool blinking detection systems), "wrap photo" (printed photo of subject held up to the camera, photo is moved back and forth to fool liveness detection systems), and video replay attacks (tablet or screen held up to the camera while playing a video of subject).

CASIA cropped [4, 5] is the customized image datasets that were created from the original CASIA images by using OpenCV Haar cascades [15]. Furthermore, we created custom resized CASIA image sets with different resolutions and lower dimensions. In our work, we considered high resolution image as 720x1280 (portrait style), standard-resolution image as 640x480 (landscape style) and 480x640 (portrait style). We also experimented on custom resized image datasets. For custom resized image datasets, we modified images on different resolutions and dimensions, such as 240x320, and 225x225. Overall, there are approximately 127,000 images that were used for this work.

## 5 Results and Discussions

The training pattern of the CNNs varies depending on time and memory availability of CPU/GPUs. To achieve the realistic performances, we ran different implementations of CNNs, including the 'modified ResNet', on all the datasets separately for multiple times and observed their performances and minute variations. In Table 1, we compare the test accuracies of the 'modified ResNet' to the DWT implemented 'modified ResNet' on ATVS Iris [2, 3], CASIA Two Class [4, 5], and CASIA cropped [4, 5] datasets. Also, we compared the test accuracies on our previously proposed and implemented 'modified VGGNet' [16] and with DWT implemented 'modified VGG Net' [16] with the above-mentioned datasets. The best results we achieved using the DWT implemented 'modified ResNet' on different segments of image datasets, with the quickest execution time compared to the others, are shown in Table 1.

The quickest execution speed achieved through the combined implementation on DWT and Modified ResNet, and the test accuracy is in the high-end acceptable range. By using the same technique, we obtained the Receiver Operating Characteristic curve (ROC) for the ATVS Iris datasets with 98% True Positive Rate as shown in Fig 5(a). The average error rate for our proposed model is in the range of (3.6% – 5.2 %).

| Classification Techniques Used | Datasets Used | | | Average Execution Time for ATVS Iris (in seconds) |
|---|---|---|---|---|
| | ATVS Iris | CASIA Two Class | CASIA Cropped | |
| Modified ResNet [34 layers] | 94.40 | 91.0 | 85.70 | 1,962 |
| DWT + 'Modified ResNet' [34 layers] | 92.57 | 90.80 | 82.4 | 1311 |
| 'Modified VGG Net' [19 layers] *[ref. 16]* | 97.00 | 96.00 | 95.00 | 1459 |
| DWT + 'Modified VGG Net' [19 layers] *[ref. 16]* | 89.00 | 86.00 | 78.00 | 3045 |

Table 1. Classification accuracies and average execution time of different Classification Techniques implemented

The test accuracy reported of the 'modified ResNet' was significant, irrespective of its very deep network architecture. The training and validation process is reasonably fast by maintaining the skip connection techniques at initial stages of training and expanding all the 32 convolution layers towards the end of training for fast training. Our 'modified ResNet' uses optimum CPU/GPU memory. The 'modified ResNet' efficiently handled standard black and white ATVS iris image [2, 3] and CASIA two-class, cropped, and custom resized face images datasets [4, 5] with precision.

The ROC curve shown in Fig 5(a), explains the True Positive Rate (TPR) versus the False Positive Rate (FPR) for our proposed 'modified ResNet' on different image datasets. In the best-case scenario, TPR should be as close as possible to 1.0, meaning that during training and validation none of the images were rejected by mistake. The FPR should be as close to 0.0 as possible, meaning that all presentation attacks were rejected, and none were mistakenly accepted as real users' image. In all cases, the area under the curve remains in the range of 0.96 to 1.00, which is close to the ideal value of 1.0. However, the TPR is generally higher for the controlled grayscale ATVS image datasets compared to less controlled CASIA color image datasets and with CASIA cropped datasets.

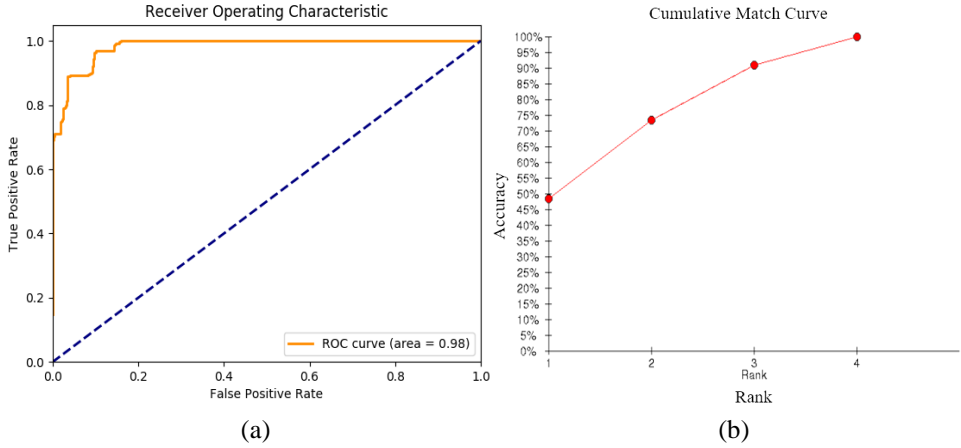

Fig5 (a). Receiver Operating Characteristic Curve and (b). Cumulative match Characteristics (CMC) Curve got on binary classification on ATVS dataset using DWT + 'Modified ResNet' classification technique.

We achieved the 90.1% rank 3 identification and recognition accuracy with precision as shown in Fig 5(b) in the cumulative match characteristics (CMC) curve. Although, we start getting 48.3% rank 1 identification accuracy and 73.3% rank 2 identification accuracy as mentioned in the CMC curve.

## 6      Conclusion

In this paper, we have performed an extensive analysis on how to obtain quick classification with high accuracy exploring on the popular spoofing datasets. Our proposed 'modified ResNet' in combination with DWT feature extraction technique implemented, perform promisingly fast with encouragingly high test accuracy. In the fully connected dense layer and flatten layer we implemented softmax function logic. Our future work will focus on testing our framework with other popular biometric datasets and observing stability and performance. Our future plan includes an extensive focus on the test accuracy and improving the overall performance of our proposed architecture.

## 7      Acknowledgment

This research is based upon work supported by the Science & Technology Center: Bio/Computational Evolution in Action Consortium (BEACON) and the Army Research Office (Contract No. W911NF-15-1-0524).

# 8    References

[1] K. He, X. Zhang, S. Ren, and J. Sun. "Deep Residual Learning for Image Recognition." arXiv:1512.03385v1 [cs.CV] 10 Dec 2015.

[2] J. Fierrez, J. Ortega-Garcia, D. Torre-Toledano and J. Gonzalez-Rodriguez, "BioSec baseline corpus: A multimodal biometric database", Pattern Recognition, Vol. 40, n. 4, pp. 1389-1392, Apr 2007.

[3] J. Galbally, J. Ortiz-Lopez, J. Fierrez and J. Ortega-Garcia, "Iris liveness detection based on quality related features", *In Proc. Intl. Conf. on Biometrics,* ICB, pp. 271-276, New Delhi, India, Mar 2012.

[4] Z. Zhiwei, J. Yan, S. Liu, Z. Lei, D. Yi, S.Z. Li, "A face antispoofing database with diverse attacks", *In Proc of IAPR International Conference on Biometrics* (ICB), Beijing, China, pp. 26–31, 2012.

[5] Chinese Academy of Sciences (CASIA), Institute of Automation, "Antispoofing dataset [online]", *http://biometrics.idealtest.org/dbDetailForUser.do?id=3*, 2010. http://www.cbsr.ia.ac.cn/english/FASDB_Agreement/Agreement.pdf, 2010.

[6] R. Rao. "Wavelet Transforms.", Encyclopedia of Imaging Science and Technology, Wiley Online Library. *https://doi.org/10.1002/0471443395.img112*. Jan 2002.

[7] J. Yang, Z. Lei, and S. Li., "Learn convolutional neural network for face anti-spoofing", arXiv:1408.5601v2 [cs.CV] Aug 2014.

[8] D. Menotti, G. Chiachia, A. Pinto, W. R. Schwartz, H. Pedrini, A. Falcão, and A. Rocha, "Deep representations for iris, face, and fingerprint spoofing detection", arXiv:1410.1980v3 [cs.CV], Pre-print of article that will appear in IEEE Transactions on Information Forensics and Security (T.IFS). 29 Jan 2015.

[9] I. Manjani, S. Tariyal, M. Vatsa, R. Singh, and A. Majumdar. "Detecting Silicone Mask-Based Presentation Attack via Deep Dictionary Learning." IEEE Transactions on Information Forensics and Security, vol. 12, no. 7, July 2017.

[10] A. Pentland and T. Choudhury, "Face recognition for smart environments," Computer, vol. 33, no. 2, pp. 50–55, 2000.

[11] P. J. Phillips, A. Martin, C. L. Wilson, and M. Przybocki, "An introduction to evaluating biometric systems," Computer, vol. 33, no. 2, pp.56–63, 2000.

[12] P. J. Phillips, H. Moon, S. A. Rizvi, and P. J. Rauss, "The FERET evaluation methodology for face-recognition algorithms," IEEE Trans. Pattern Analysis and Machine Intelligence, vol. 22, no. 10, pp. 1090–1104, Oct 2000.

[13] J. Daugman. "How Iris Recognition Works." IEEE Transactions on Circuits and Systems for Video Technology, vol. 14, No. 1, Jan. 2004.

[14] T. S. Lee. "Image Representation Using 2D Gabor Wavelets." IEEE Transaction on Pattern Analysis and Machine Intelligence, vol. 18, No. 10, Oct 1996.

[15]  A. Mordvintsev and Abid K. (2013). Face Detection using Haar Cascades. [online] OpenCV Tutorial. Available: *https://opencv-python-tutroals.readthedocs.io/en/latest/py_tutorials/py_objdetect/py_face_detection/py_face_detection.html.*

[16] P. Chatterjee, and K. Roy. "Anti-spoofing Approach Using Deep Convolutional Neural Network." In book: Recent Trends and Future Technology in Applied Intelligence. DOI: 10.1007/978-3-319-92058-0_72. Jan 2018.